\documentstyle[12pt]{article}
\begin{document}
\title{Diffraction at small $M^2/Q^2$ in the QCD dipole picture}
\author{A. Bialas$^{a,b},$ H. Navelet$^a$ and R. Peschanski$^a$ \\ \\
$^a$Service de Physique Theorique  CEA-Saclay \\ F-91191
Gif-sur-Yvette Cedex, France\\
$^b$M.Smoluchowski Institute of Physics, Jagellonian University\\ Reymonta
4, 30-059 Cracow, Poland}
\maketitle

\begin{abstract}
Using the QCD dipole picture of BFKL dynamics, the quasi elastic component of the diffractive $\gamma ^*$-dipole cross-section, dominating at small $M^2/Q^2,$
is calculated.
\end{abstract}

{\bf 1.} Diffraction dissociation of virtual photons, a process  being on
the border line between the perturbative and non-perturbative regions,
 represents a challenging problem for  Quantum Chromodynamics \cite{li}.
It is therefore of no surprise that it became a subject of numerous
studies by various methods \cite{r2}-\cite{mpr}.

The present investigation continues our effort \cite{bp1}-\cite{bc} to reach a
unified description
of both total and diffractive cross-sections of virtual photons in the
framework of the QCD dipole picture of high-energy interactions
\cite{mu1,mu4}. More precisely, it is and attempt to obtain -within
the approximations inherent to the QCD dipole model - an  exact formula
for the quasi-elastic component \cite{bp1} of the (virtual) photon diffractive
dissociation at arbitrary momentum transfer.
It is a generalization of a
recent calculation \cite{bc} of the forward photon diffractive processes.
The need for  an {\it exact} formula is emphasized by (i) recent success of the
approximate dipole model description \cite{ro,mpr} of the HERA data \cite{he1} and
(ii) improved accuracy of recent measurements
\cite{he} which provide already first results on momentum transfer dependence
of diffractive structure functions.

 We restrict ourselves to the so called
quasi elastic component of the
diffractive cross-section, dominating at small $M^2/Q^2$ where $M$ is the mass
of the produced diffractive system and $Q$ is the virtuality of the photon.
 The calculation
of the high mass component at non-zero momentum transfer was already
presented in \cite{bnp}.

In the QCD dipole picture  the cross-section for the quasi-elastic scattering
of a virtual photon on a single dipole target can be written as \cite{bp1}
\begin{eqnarray}
\frac{d\sigma}{dM^2 d^2p_T} = \frac {2N_c}{4\pi^2}\int d^2k
|<\vec{k},M^2|T^{qel}_{p_T}|Q>|^2 \label{2}
\end{eqnarray}
where $p_T$ is the transverse momentum after the collision,
 $\vec{k} =(|k|,\phi_k)$ is
the relative tranverse momentum in the ($q\bar{q}$) pair, and
\begin{eqnarray}
<\vec{k},M^2|T^{qel}_{p_T}|Q> = \int_0^1 \int d^2r<\vec{k},M^2|\vec{r},z>
<\vec{r},z|T_{p_T}| \vec{r},z> \Psi(\vec{r},z;Q).   \label{a2}
\end{eqnarray}
Here $\vec{r}=(|r|,\phi_r)$ is the relative transverse distance in the
($q\bar{q}$) pair,
 $z$ is the light-cone momentum fraction of  one of the quarks and
$<\vec{r},z|T_{p_T}| \vec{r},z>$ is the elastic dipole amplitude for scattering
of the dipole of tranverse size $r$ on the target dipole of size $r_0$.
$\Psi(\vec{r},z;Q)$ are the light-cone photon wave functions \cite{bks}.

Following the argument of \cite{bp1} (c.f. also \cite{bc}), Eq.(\ref{2})
 can be rewritten in the form
\begin{eqnarray}
\frac{d\sigma}{dM^2 d^2p_T} = \frac {2N_c}{4\pi^2}
\int_0^1 dz \ z(1-z) \ \frac{1}2  d\phi_k \left| G(\hat M, z,p_T;x_{\cal
P})\right|^2 .                        \label {a2a}
\end{eqnarray}
where
\begin{equation}
 G(k,z,p_T;x_{\cal P}) =
\frac 1{2\pi}\int d^2 r \ e^{i \hat M \dot r}\
 <\vec{r},z|T_{p_T}| \vec{r},z>
\Psi (r,z;\hat Q) \label{3b}
\end{equation}
and   $\hat{M}$  is the vector   of length
\begin{equation}
\hat{M}= M\sqrt{z(1-z)}                \label{2a}
\end{equation}
parallel to $\vec{k}$.
Eq.(\ref{3b}) is the starting point of our calculation.

{\bf 2.} The difficult  part of the task is to perform the integration over
$d^2r$ in (\ref{3b}). To this end we need an adequate formula for
 the dipole-dipole
amplitude $<\vec{r},z|T_{p_T}| \vec{r},z>$. For $p_T=0$ an explicit formula is
available \cite{mu1} and this allowed the calculation of \cite{bc}. For
$p_T \neq 0$, however, the calculation is much more involved. In the
present paper we perform it using the methods
 developed  recently in \cite{np1}.

The starting point is the general expression \cite{l1}
\begin{eqnarray}
<\vec{r},z|T_{p_T}| \vec{r},z> = 4\pi \alpha^2
 \sum_{n=-\infty}^{n=\infty}\int \frac {d\nu}{\pi} \
e^{\omega(n,\nu)Y} d_{n,\nu}\  |r| E^{n,\nu}_{p_T}(r)
|r_0| \bar{E}^{n,\nu}_{p_T}(r_0),  \label{4}
\end{eqnarray}
where
\begin{equation}
d_{n,\nu}=\left(16\left[\nu^2+\left(\frac{n-1}2\right)^2\right]
\left[\nu^2+\left(\frac{n+1}2\right)^2\right]\right)^{-1} , \label{7}
\end{equation}
\begin{equation}
\omega(n,\nu)= \frac{2\alpha N_c}{\pi} \left(\psi(1)-Re
[\psi(\frac12(|n|+1)+i\nu)]\right)   \label{4a}
\end{equation}
and $E^{n,\nu}_{p_T}(r)$ are   eigenfunctions of the conformal operator in
the
mixed representation defined in  \cite{l1}. The explicit expression for
$E^{n,\nu}_{p_T}(r)$  is fairly complicated \cite{np1,l1} but, fortunately, we
shall not need
it here. We only quote, for future reference, the formula for $E^{n,\nu}_0(r)$:
\begin{equation}
E^{n,\nu}_0(r) =   |r|^{-2i\nu} e^{in\phi_r}\ . 
\label{7b}
\end{equation}
 The sum over $n$ in (\ref{4})
was introduced
for technical reasons. In the high-energy (large $Y$) limit, the term with
$n=0$ dominates.

To proceed,  it is useful at this point  to  introduce the explicit formulae
for $\Psi(\vec{r},z;Q)$. We  write them in the form
\begin{equation}
\Psi(\vec{r},z;Q) =C \hat{Q} \ \Phi_{T,L}(z)\chi_{T,L}(r,\hat{Q})
\label{3a}
\end{equation}
with  $\hat{Q} = Q\sqrt{z(1-z)}$, $C= \sqrt{\alpha_{em}}e_{(f)}/2\pi$
($e_{(f)}$ is the charge of a quark) and
\begin{equation}
 \Phi_T= z;\;\;\;\;\Phi_L= 2\sqrt{z(1-z)};\;\;\;\;
 \chi_T= e^{i\phi_r}K_1(\hat{Q}r)
;\;\;\;\;  \chi_L= K_0(\hat{Q}r) \ . \label{a3a}
\end{equation}
The subscripts $(T,L)$ denote transverse (right-handed) and longitudinal
polarizations of the incident photon. Using (\ref{3a})
one  obtains
\begin{equation}
 G_{T,L}(k,z,p_T;x_{\cal P}) =D\ |r_0|\hat{Q} \Phi_{T,L}(z)\ \sum_n\int
\frac
{d\nu}{\pi}
\ d_{n,\nu} \ e^{\omega(n,\nu)Y}  E^{n,\nu}_{p_T} (r_0)
  g_{T,L}^{n,\nu},\label{3}
\end{equation}
where $D\equiv 4\pi\alpha^2 C= 2\alpha^2 \sqrt{\alpha_{em}}e_{(f)}$,
\begin{equation}
g_{T,L}^{n,\nu}(\hat{Q},\hat{M})=\frac 1{2\pi}\int
d^2r\psi_{T,L}(r,\hat{Q},\hat{M})
E^{n,\nu}_{p_T} (r)    \label{a4a}
\end{equation}
and
\begin{equation}
\psi_{T,L}(r,\hat{Q},\hat{M})= e^{i
\hat{ M} r}\ \chi_{T,L}(r,\hat{Q}) |r|\ .   \label{4b}
\end{equation}
We can now expand $\psi_{T,L}(r,\hat{Q},\hat{M})$ in terms of
conformal eigenfunctions at $p_T=0$:
\begin{equation}
\psi_{T,L}(r,\hat{Q},\hat{M})= \sum_{n}\int
d\nu\ \psi^{n,\nu}_{T,L}(\hat{Q},\hat{M})
\frac{E^{n,\nu}_0 (r)}{|r|^2}
  \label{4c}
\end{equation}
The inverse tranform is
\begin{equation}
\psi^{n,\nu}_{T,L}(\hat{Q},\hat{M})=\frac{1}{2\pi^2} \int d^2r\ 
\psi_{T,L}(r,\hat{Q},\hat{M}) E^{n,\nu}_0 (r)       \label{4d}
\end{equation}
and thus we obtain
\begin{equation}
g_{T,L}^{n,\nu} (\hat{Q},\hat{M})= \frac1{2\pi} \sum_{n^{\prime}}\int d\nu^{\prime}
\psi^{n^{\prime},\nu^{\prime}}_{T,L}(\hat{Q},\hat{M})
\  {\cal I}^{n,n^{\prime},\nu,\nu^{\prime}}_{p_T}.   \label{4e}
\end{equation}
We see that the integration over $r$ is reduced to
\begin{equation}
{\cal I}^{n,n^{\prime},\nu,\nu^{\prime}}_{p_T}\equiv
 \int d^2r \frac
 {E^{n,\nu}_{p_T} (r) E^{n^{\prime},\nu^{\prime}}_0 (r)}{|r|^2}
\label{a4d}
\end{equation}
which was explicitly calculated in \cite{np1} with the result
\begin{eqnarray}
 {\cal I}^{n,n^{\prime},\nu,\nu^{\prime}}_{p_T}
 &=& \frac {\pi}2 (-1)^{\frac {n-n ^{\prime }}2}\left[\frac {p_T}8\right]^{\tilde {\mu}-\tilde {\mu}^{\prime}}\
\left[\frac {\bar {p_T}}8\right]^{ {\mu}- {\mu}^{\prime}} \nonumber \\ &\times&
\frac {\Gamma (1\!-\! \mu)}{\Gamma (\tilde \mu)}\
\frac{\Gamma \left(\frac { {\mu}+ {\mu}^{\prime}}2\right)
\Gamma \left(\frac {- {\mu}+ {\mu}^{\prime}}2\right)}
{\Gamma \left( 1\!-\!\frac {\tilde {\mu}+\tilde {\mu}^{\prime}}2\right)
\Gamma \left( 1\!-\!\frac{\tilde {\mu}^{\prime}-\tilde {\mu}}2\right)}\ ,
\label{14}
\end{eqnarray}
for $n-n'$ even and $0$ for  $n-n'$  odd. Here
$ \mu = i\nu -\frac n2\ ;\tilde\mu = i\nu +\frac n2\ .$

We can  now calculate $\psi^{n,\nu}(\hat{Q},\hat{M})$. From (\ref{4d}) and
(\ref{7b}) we  have
\begin{equation}
\psi_{T,L}^{n,\nu}(\hat Q,\hat M) = \frac{1}{2\pi^2} \int \ d^2 r
 \ \psi_{T,L}(r,\hat Q,\hat M) \mid r\mid^{-2i\nu}\ e^{in\phi_r},
\label{15}
\end{equation}
where $\phi_r$ is the $r$ azimuthal angle with respect to $p_T.$

Using
\begin{equation}
 \int d\Psi \ e^{i\hat M \rho \cos \Psi \pm im \Psi} \equiv 2\pi\
e^{im\frac{\pi}2}
\ J_m(\hat M\rho),
\label{16}
\end{equation}
we find for right-handed photons (for  left-handed photons, $n+1$ should be replaced by $n-1$):
\begin{equation}
\psi_{T}^{n,\nu}(\hat Q,\hat M) = \frac1{\pi} \int \ r^2 dr   r^{-2i\nu}
e^{i(n+1)\frac{\pi}2-i\phi_k} J_{n+1}(\hat Mr)\ K_1(\hat Q r)\ .
\label{17}
\end{equation}
For the longitudinal photons we obtain
\begin{equation}
\psi_{L}^{n,\nu}(\hat Q,\hat M) = \frac1{\pi} \int r^2 dr  \  r^{-2i\nu}
e^{in\frac{\pi}2} J_{n}(\hat Mr)\ K_0(\hat Q r)
\label{18}
\end{equation}
Using now \cite{gr}
\begin{eqnarray}
\int_0^{\infty} d\rho\ \rho^{-\lambda}\ K_h(\hat Q\rho)\ J_l(\hat M \rho) =
\frac 14 \left(\frac MQ\right)^l \left(\frac {\hat Q}2\right)^{\lambda-1}\ \times
\nonumber \\
\frac {\Gamma\left(\frac {1-\lambda+l+h}2\right)\
 \Gamma\left(\frac {1-\lambda+l-h}2\right)}{\Gamma(l+1)}\
 {_2F_1}\left(\frac {1\!-\!\lambda\!+\!l\!+\!h}2,
\frac {1\!-\!\lambda\!+\!l\!-\!h}2,l\!+\!1;-\frac {M^2}{Q^2}\right)
  \label{18a}
\end{eqnarray}
and
\begin{eqnarray}
 {_2F_1}\left(\frac {1-\lambda+l+h}2,
\frac {1-\lambda+l-h}2,l+1;-\frac {M^2}{Q^2}\right) = \nonumber \\
 = (\beta)^{\frac {1-\lambda+l- h}2}\ {_2F_1}\left(\frac {1-\lambda+l-h}2,
\frac {1+\lambda+l-h}2,l+1;1-\beta\right)
\label{19}
\end{eqnarray}
where $h,l$ are positive integers, we find
\begin{eqnarray}
\psi_{T,right}^{n,\nu}(\hat Q,\hat M) = \frac1{4\pi}
e^{-i\pi\frac{n+1}2-i\phi_k}\
 \left(\frac MQ\right)^{\mid n+1\mid}\ \left(\frac {\hat Q }2\right)^{2i\nu-3}
\ \times \nonumber \\
 \frac {\Gamma\left(-i\nu+2+\frac {\mid n+1\mid}2\right)\
\Gamma\left(-i\nu+1+\frac {\mid n+1\mid}2\right)}{\Gamma(\mid n+1\mid+1)}\
(\beta)^{-i\nu+1+\frac {\mid n+1\mid}2}\ \times \nonumber \\
 \ {_2F_1}\left(-i\nu+1+\frac {\mid n+1\mid}2,
\frac {\mid n+1\mid}2+i\nu-1,\mid n+1\mid+1;1-\beta\right), \label{20}
\end{eqnarray}
\begin{eqnarray}
\psi_{L}^{n,\nu}(\hat Q,\hat M) = \frac1{4\pi} e^{-i\pi\frac{n}2-i\phi_k}\
 \left(\frac MQ\right)^{\mid n\mid}\ \left(\frac {\hat Q }2\right)^{2i\nu-3}
\ \times \nonumber \\
 \frac {\Gamma^2\left(-i\nu+\frac32+\frac {\mid n\mid}2\right)}{\Gamma(\mid n\mid+1)}\
 (\beta)^{-i\nu+\frac 32+\frac {\mid n\mid}2}\ \times\nonumber \\{_2F_1}
\left(-i\nu+\frac 32+\frac {\mid n\mid}2,\frac {\mid n
\mid}2+i\nu-\frac12,\mid n\mid+1;1-\beta\right).
\label{21}
\end{eqnarray}
Next we denote
\begin{eqnarray}
r_{T}\left(\nu_1,\nu_2\right) = 2 B\left(-i\nu_2+i\nu_1+2,-i\nu_2+i\nu_1\right)\
\nonumber \\ r_{L}\left(\nu_1,\nu_2\right) =
4 B\left(-i\nu_2+i\nu_1 +1,-i\nu_2+i\nu_1 + 1\right),
\label{23}
\end{eqnarray}
which are factors coming from integrals over the variable $z.$
All in all,  we obtain
\begin{eqnarray}
\frac{d\sigma}{dM^2 d^2p_T} &=&\frac{N_c}{2 \pi} \int_0^1 dz \ z(1-z)\ \left| G(\hat M, z;x_{\cal P})\right|^2 \nonumber \\
&=& \frac{N_c}{8 \pi} Q^2 D^2 \sum_{n_1} \int \frac {d\nu_1}{\pi}\
e^{ \omega\left(n_1,\nu_1\right)\ Y}
  d_{n_1,\nu_1} \ \mid r_0\mid  \bar E^{n_1,\nu_1}_{p_T}(r_0)\nonumber \\
&\times& \sum_{n_2} \int \frac {d\nu_2}{\pi}\
 e^{ \omega\left(n_2,\nu_2\right)\ Y}
   \bar d_{n_2,\nu_2} \ \mid r_0\mid
 E^{n_2,\nu_2}_{p_T}(r_0)\nonumber \\
&\times&  \sum_{n'_1} \int \frac {d\nu'_1}{\pi}\ \psi_{T,L}^{n'_1,\nu'_1}(Q,M)\
 {\cal I}^{n_1,n'_1,\nu_1,\nu'_1}_{P_T}
\nonumber \\
 &\times& \sum_{n'_2} \int \frac {d\nu'_2}{\pi}\
\bar \psi_{T,L}^{n'_2,\nu'_2}(Q,M)\ \bar {\cal I}^{n_2,n'_2,\nu_2,\nu'_2}_{P_T}
\ r_{T,L}\left(\nu'_1,\nu'_2\right).
\label {24}
\end{eqnarray}
This completes the calculation.

Note that at $p_T=0,$ we have
\begin{equation}
I^{nn',\nu\nu'}_0= 2\pi^2\delta_{nn'}\delta(\nu-\nu')   \label{24a}
\end{equation}
and thus, using (\ref{7b}), we obtain

\begin{eqnarray}
\frac{d\sigma}{dM^2 d^2p_T}\mid_{p_T=0}= 8 \pi \alpha_{em}\alpha^4 N_c e^2_f \left(\frac {Q r_0}2\right)^2 \nonumber \\
\sum_{n_1} \int
\frac
{d\nu_1}{\pi}\ e^{ \omega\left(n_1,\nu_1\right)\ Y}
        d_{n_1,\nu_1}  (r_0)^{2i\nu_1}
 \psi_{T,L}^{n_1,\nu_1}(Q,M)\ \nonumber \\
\sum_{n_2} \int \frac {d\nu_2}{\pi}\
 e^{ \omega\left(n_2,\nu_2\right)\ Y}
\bar d_{n_2,\nu_2}   (r_0)^{-2i\nu_2}\bar \psi_{T,L}^{n_2,\nu_2}(Q,M)\
 \ r_{T,L}\left(\nu_1,\nu_2\right),           \label {25}
\end{eqnarray}
which, for $n_1=n_2=0,$ is identical to the result obtained in \cite{bc}.

{\bf 3.}
To obtain more insight into the $P_T$ dependence of the cross-section given by 
(\ref{24}), it is useful to evaluate the integrals over $\nu'_1$ and $\nu'_2$ in terms of residues of the relevant poles of the integrands. To this end we observe that the convergence properties of these integrals are determined by the factor
\begin{equation}
\left(\frac 1{\hat p_T}\right)^{2i\nu'} \equiv \left(\frac {4Q}{P_T\sqrt{\beta}}\right)^{2i\nu'}.
\label{26}
\end{equation}
This factor is easily identified when the explicit expressions (\ref {14},\ref {17},\ref {18}) are introduced into the product $\Psi^{n',\nu'}_{T,L}(Q,M)\ {\cal I}^{n,n',\nu,\nu'}_{P_T}.$
Thus for 
\begin{equation}
\hat p_T \equiv \frac {P_T\sqrt{\beta}}{4Q} < 1,
\label{27}
\end{equation}
the contour integrals in $\nu'_1$ and $\nu'_2$ must be choosen in the lower half of the complex plane, so only the poles at $\Re (i\nu') \le 0$ contribute.  
 The integrals are given by the residues of the {\it moving} poles $i\nu'=\pm i\nu -2p$ leading to an expansion  in terms of powers $\left(\hat p_T\right)^{4p}$
 of the ``reduced'' transverse momentum(\ref{27}). Thus one obtains:
\begin{eqnarray}
\frac{d\sigma}{dM^2 d^2P_T}= 8 \pi \alpha_{em}\alpha^4 N_c e^2_f \left(\frac {Q r_0}2\right)^2\ \times
 \nonumber \\
 \sum_{n_1} \int \frac {d\nu_1}{\pi}\
 e^{ \omega\left(n_1,\nu_1\right)\ Y}
     d_{n_1,\nu_1}  \bar  E^{n_1,\nu_1}_{P_T}(r_0) \psi_{T,L}^{n_1,\nu_1}(Q,M)\  \sum_{n_2} \int \frac {d\nu_2}{\pi}\ \times \nonumber \\
e^{ \omega\left(n_2,\nu_2\right)\ Y}           \bar d_{n_2,\nu_2}  E^{n_2,\nu_2}_{P_T}(r_0)\bar \psi_{T,L}^{n_2,\nu_2}(Q,M)\ \ r_{T,L}\left(\nu_1,\nu_2\right)\times \left[1 + {\cal O} \left(\hat p_T\right)^4\right].
\label {28}
\end{eqnarray}

On the other hand, for 
\begin{equation}
\hat p_T \equiv \frac {P_T\sqrt{\beta}}{4Q} > 1,
\label{29}
\end{equation}
the contour integrals must be closed on the upper half of the complex plane. Consequently, only poles of the $\Psi$ functions at $\Re (i\nu')>0$ contribute. The integrals are thus given by {\it fixed} poles at $i\nu'=\frac 32 +2p$ and $i\nu'=\frac 52 +2p$ leading to an expansion independent of $\nu,$ resulting in a series expansion in powers of $ 1 /{\hat p_T}$.
The  two regimes are thus governed by different types of singularities, as was already noticed about vertices of BFKL pomerons \cite {np1}.

{\bf 4.} We would like to close this note by the following remarks.

(a) The presented calculation extends the results of \cite {bc} to non-vanishing momentum transfer and thus -together with \cite {bnp} where the large mass component was calculated- it completes the derivation of the hard diffractive cross-section in the QCD dipole picture.

(b) The $P_T$-dependence in Eqs. (\ref {24}) and (\ref {28}) may be modified by the proton form factor and possibly other non-perturbative effects. As the same effects would operate also for the large mass component in \cite {bnp}, one may hope that the relative weight of the two components is reasonably well described by our result. A comparison with data should therefore be a significant test of  the dipole approach. It would be particularly interesting to compare this new result with the satisfactory description of data obtained in an approximate version of the dipole model \cite {mpr}. 

c) Our formulae (\ref{24},\ref{28}) sum over all conformal spin values $n$ allowed by selection rules of the BFKL vertices \cite {np1}. At high energies the term with $n=0$ -corresponding to the so-called ``hard pomeron'' \cite {li}- dominates. Consequently the phenomenological discussion is usually	restricted to $n=0.$ As argued recently \cite {mp}, however, the analysis of total cross-sections data \cite {dl1} allows the interpretation of the next term in the expansion, namely $n=2,$ as being at the origin of the so-called ``soft pomeron'' \cite{dl2}. Our formula can thus provide an independent test of this hypothesis in hard diffraction dissociation.

To summarize, using the QCD dipole picture of BFKL dynamics we have calculated the quasi elastic component of the diffractive $\gamma ^*$-dipole cross-section, dominating at small $M^2/Q^2.$ This work, together with \cite {bnp}, completes the calculation of hard diffraction in the framework  of the  QCD dipole model and thus  allows an extension of the previous phenomenological analyses based on approximate formulae \cite {ro,mpr}. This should in turn allow to separate perturbative and non perturbative QCD contributions to the diffraction on the proton target. Finally,
let us note that the formulae obtained in \cite {bnp} and here can easily  be generalized to $\gamma ^*- \gamma ^*$ interactions and thus provide a basis for 
a phenomenological analysis of future collider data.

{\bf Acknowledgements}

\noindent We thank S.Munier and Ch.Royon for fruitful discussions. A.B. thanks the Service de Physique Th\'eorique of Saclay for the kind hospitality. This work was supported in part by the KBN Grant No 2 P03B086 14.

\end{document}